\documentclass[preprint,aps,nofootinbib]{revtex4}
\usepackage{graphicx}
\usepackage{epstopdf}
\usepackage{subfigure}
\usepackage{textcomp}
\def\bkR{{\rm I\kern-.17em R}}
\def\bkC{{\rm \kern.24em \vrule width.05em height1.4ex depth-.05ex \kern-.26em C}}

\def\PR{{\it Phys. Rev.} }

\def\be{\beta}

\def\frac#1#2{{\textstyle{{#1}\over {#2}}}}

\def\lsim{\mathrel{\rlap{\lower4pt\hbox{\hskip1pt$\sim$}}
    \raise1pt\hbox{$<$}}}
\def\gsim{\mathrel{\rlap{\lower4pt\hbox{\hskip1pt$\sim$}}
    \raise1pt\hbox{$>$}}}
\def\sqr#1#2{{\vcenter{\vbox{\hrule height.#2pt
         \hbox{\vrule width.#2pt height#1pt \kern#1pt
         \vrule width.#2pt}
         \hrule height.#2pt}}}}

\def\laq{\raise 0.4 ex \hbox{$<$}\kern -0.8 em\lower 0.62 ex\hbox{$\sim$}}
\def\gaq{\raise 0.4 ex \hbox{$>$}\kern -0.7 em\lower 0.62 ex\hbox{$\sim$}}

\def\be{\begin{equation}}
\def\ee{\end{equation}}
\def\beqa{\begin{eqnarray}}
\def\eeqa{\end{eqnarray}}

\def\dalemb#1#2{{\vbox{\hrule height.#2pt
        \hbox{\vrule width.#2pt height#1pt \kern#1pt \vrule width.#2pt}
        \hrule height.#2pt}}}

\def\dalemb#1#2{{\vbox{\hrule height.#2pt
        \hbox{\vrule width.#2pt height#1pt \kern#1pt \vrule width.#2pt}
        \hrule height.#2pt}}}

\def\gtorder{\mathrel{\raise.3ex\hbox{$>$}\mkern-14mu
             \lower0.6ex\hbox{$\sim$}}}
\def\ltorder{\mathrel{\raise.3ex\hbox{$<$}\mkern-14mu
             \lower0.6ex\hbox{$\sim$}}}

\begin{document}

\rightline{March 2011}

\title{Entropic Gravity, Phase-Space Noncommutativity and the Equivalence Principle}

\author{Catarina Bastos\footnote{E-mail: catarina.bastos@ist.utl.pt}}

\vskip 0.3cm

\affiliation{Instituto de Plasmas e Fus\~ao Nuclear, Instituto Superior T\'ecnico \\
Avenida Rovisco Pais 1, 1049-001 Lisboa, Portugal}

\author{Orfeu Bertolami\footnote{Also at Instituto de Plasmas e Fus\~ao Nuclear,
Instituto Superior T\'ecnico, Avenida Rovisco Pais 1, 1049-001 Lisboa, Portugal. E-mail: orfeu.bertolami@fc.up.pt}}

\vskip 0.3cm

\affiliation{Departamento de F\'\i sica e Astronomia, Faculdade de Ci\^encias da Universidade do Porto \\
Rua do Campo Alegre 687, 4169-007 Porto, Portugal}

\author{Nuno Costa Dias\footnote{Also at Grupo de F\'{\i}sica Matem\'atica, UL,
Avenida Prof. Gama Pinto 2, 1649-003, Lisboa, Portugal. E-mail: ncdias@meo.pt}, Jo\~ao Nuno Prata\footnote{Also at Grupo de F\'{\i}sica Matem\'atica, UL,
Avenida Prof. Gama Pinto 2, 1649-003, Lisboa, Portugal. E-mail: joao.prata@mail.telepac.pt}}


\affiliation{Departamento de Matem\'{a}tica, Universidade Lus\'ofona de
Humanidades e Tecnologias \\
Avenida Campo Grande, 376, 1749-024 Lisboa, Portugal}


\vskip 0.5cm

\begin{abstract}

\vskip 0.5cm

{We generalize E. Verlinde's entropic gravity reasoning to a phase-space noncommutativity set-up. This allows us to impose a bound on the product of the noncommutative parameters based on the Equivalence Principle. The key feature
of our analysis is an effective Planck's constant that naturally arises when accounting for the noncommutative features of the phase-space.}

\end{abstract}

\maketitle

\section{Introduction}

Recently, it has been argued that Newton's inverse square law (ISL) can be derived from thermodynamical considerations and from the Holographic Principle \cite{Verlinde}. In this formulation the ISL is obtained from somewhat more fundamental
entities such as energy, entropy, temperature and the counting of degrees of freedom which is set by the Holographic Principle \cite{tHooft,Susskind}. Thus, the key element in the derivation of gravity, as in the case of black holes, is information and its relationship with entropy. The variation of entropy is assumed to be generated by the displacement of matter, leading to an entropic force, which takes the form of gravity. Another interesting relationship arising from this derivation is that the force is proportional to the temperature, and the relevant fundamental constants are chosen in order to match Newton's second law of mechanics, $F=ma$, once the Rindler-Unruh temperature for accelerated frames is assumed.

An immediate implication of this entropic derivation of the ISL is that gravity is not a fundamental interaction, or at least, it is not directly related with the spin-2 state associated with the graviton, which is found in a fundamental theory such as string theory. Actually, the arguments of Ref. \cite{Verlinde} are in line with ideas that appeared earlier, and which advocate gravity as an emerging phenomenon of a thermodynamical nature \cite{Jacobson,Padmanabhan}. Nevertheless, in Ref. \cite{Verlinde}, the fundamental standing of string theory is assumed, in as much as it is in the context of this theory that one can make sense of the Holographic Principle \cite{Maldacena,Bousso}.
Furthermore, we argue that string theory provides another basic underlying ingredient, namely an expected noncommutative structure of space-time \cite{Connes,Seiberg}.

The role that configuration space noncommutativity might play in the derivation of entropic gravity has been considered in Ref. \cite{Nicolini}. However, it has been argued elsewhere that phase-space noncommutativity is crucial in addressing problems in quantum cosmology \cite{Bastos1,Bastos5}, the Schwarzschild black hole (BH) thermodynamics \cite{Bastos2} and the BH singularity problem \cite{Bastos3,Bastos3a}. Hence, it seems just natural that this more general form of noncommutativity is considered in the context of an entropic derivation of the ISL.

Henceforth, we assume that space-time is ruled by a geometry based on a canonical phase-space noncommutative algebra. In two dimensions such a theory was studied in the context of the gravitational quantum well \cite{Bertolami1,Bertolami2}. But one can easily consider the following generalization to $d$ space dimensions:
\be\label{eq1.1}
\left[\hat q'_i, \hat q'_j \right] = i\theta_{ij} \hspace{0.2 cm}, \hspace{0.2 cm} \left[\hat q'_i, \hat p'_j \right] = i \hbar \delta_{ij} \hspace{0.2 cm},
\hspace{0.2 cm} \left[\hat p'_i, \hat p'_j \right] = i \eta_{ij} \hspace{0.2 cm}, \hspace{0.2 cm} i,j= 1, ... ,d
\ee
where $\eta_{ij}$ and $\theta_{ij}$ are antisymmetric real constant ($d \times d$) matrices and $\delta_{ij}$ is the identity matrix. This extended algebra is related to the standard Heisenberg-Weyl algebra
\be\label{eq1.2}
\left[\hat q_i, \hat q_j \right] = 0 \hspace{0.2 cm}, \hspace{0.2 cm} \left[\hat q_i, \hat p_j \right]
= i \hbar \delta_{ij} \hspace{0.2 cm}, \hspace{0.2 cm} \left[\hat p_i, \hat p_j \right] = 0 \hspace{0.2 cm},
\hspace{0.2 cm} i,j= 1, ... ,d ~,
\ee
by a class of linear non-canonical transformations. In the mathematics literature this mapping between variables is referred to as Darboux map, while in physics, the designation of Seiberg-Witten map is more often employed. These transformations are not unique. In fact, the composition of a Darboux map and a symplectic (canonical) transformation yields an equally valid Darboux map. But it is important to point out that all physical predictions (expectation values, eigenvalues, probabilities) are independent of the particular Darboux map chosen \cite{Bastos4,Bastos6}.

In what follows we evaluate the noncommutative correction to the entropic force. Our approach consists of generalizing some of the formulae leading to the ISL, when we assume an underlying noncommutative phase-space. As we will argue, this just amounts to replacing in E. Verlinde's result Planck's constant by an effective Planck's constant. Other modifications associated with the number of degrees of freedom necessarily lead to corrections to the ISL, which are in conflict with the observations. The result obtained will allow us to use the Equivalence Principle to set a bound on the noncommutativity anisotropy and, under assumptions, on the product of the noncommutative parameters. Rather remarkably, these bounds can be set without any assumption of the magnitude of the configuration space noncommutativity, in opposition to the one obtained, for instance, in the context of the noncommutative gravitational quantum well \cite{Bertolami1}.

This work is organized as follows. In the next section we review the main concepts lying at the heart of the entropic gravity approach and derive its phase-space noncommutative extension. In section 3, we use the Equivalence Principle to obtain a bound for the noncommutative parameters. Finally,  in section 4, we present our conclusions, discuss our result as well as some of its implications.

\section{Entropic gravity and phase-space noncommutativity}

Let us first review the basic features of the entropic derivation of the ISL. Consider a small piece of an holographic screen and a particle of mass, $m$, that travels at a distance $\Delta x$ from the side of the screen where space has already
emerged. The change of entropy, related to the information on the boundary, is assumed to be linear in the displacement $\Delta x$,
\be\label{Eq2.1}
\Delta S=2\pi k_B {\Delta x \over \lambda _c}~,
\ee
where $\lambda_c= \hbar/mc$ is the Compton wavelength and $k_B$ Boltzmann's constant. This result arises from saturation of the Bekenstein bound \cite{Bekenstein}, $\Delta S \le 2 \pi k_B E \Delta x / \hbar c$, $E$ being the relativistic energy. 

Consider now a mass $M$ and assume that its energy $E=Mc^2$ is projected onto a spherical holographic screen at a distance $r$ from $M$. The screen has area $A= 4 \pi r^2$ and it is divided in $N$ cells of the fundamental unit of area. This quantum of area $a_0$ is presumably associated with the square of Planck's length, $L_P^2=G \hbar/c^3$, $G$ being Newton's constant. That is, $a_0=L_P^2=G\hbar/c^3$. Thus, $N$ is given by:
\be\label{Eq2.1a}
N={A\over a_0}={{4\pi r^2} \over a_0}~.
\ee
This is an application of the Holographic Principle, according to which the number of degrees of freedom grows with the area and that the interior of some spaces (the negatively curved ones \cite{Maldacena}) can be described in terms of a theory at the boundary of the space.

The entropic force, that is, the force that arises from the Second Law of Thermodynamics, is evaluated as
\be\label{Eq2.2}
F \Delta x=T \Delta S~.
\ee
We clearly see here that this force is proportional to the temperature. The energy in the surface is given by:
\be\label{Eq2.4}
U=Mc^2~.
\ee

Considering that the surface is in thermal equilibrium at the temperature T, then all the N bits have the same probability. Thus, the energy of the surface is equipartitioned among them, as
\be\label{Eq2.5}
U={1\over2}Nk_B T~.
\ee
Finally, substituting Eqs. (\ref{Eq2.1}), (\ref{Eq2.1a}), (\ref{Eq2.4}) and (\ref{Eq2.5}) into Eq. (\ref{Eq2.2}) we obtain Newton's law for gravity, $F=GMm/r^2$.

Now, suppose that our surface is described by noncommutative geometry. We expect the underlying unit cell on the holographic screen to be modified. Let us provisionally, use the following Ansatz
\be\label{Eq2.6}
a_0 \longrightarrow a_{NC}={G \hbar_{eff}\over c^3}~,
\ee
Here $\hbar_{eff}$ is an effective Planck constant, which accounts for the noncommutative effects. Consequently, the temperature changes to
\be\label{Eq2.7}
T={G\hbar_{eff}\over ck_B}{M\over 2\pi r^2}~.
\ee
As the Compton wavelength does not correspond to an area, we shall assume that it remains unaltered by the onset of noncommutativity and similarly
\be\label{Eq2.9}
\Delta S_{NC} = \Delta S= 2 \pi k_B {mc\over\hbar} \Delta x~.
\ee
Thus, a simple calculation shows that the noncommutative correction to the force ($F_{NC}$) is such that
\be\label{Eq2.10}
{F_{NC} \over F}={\hbar_{eff} \over \hbar}~.
\ee
Similar conclusions could be reached by the following reasoning. We first consider the replacement of Planck's constant in Eq. (\ref{Eq2.6}). Indeed, it seems reasonable to expect that the quantum of area is affected by the ``fuzziness" entailed by noncommutativity. We could equally expect a further effect due to the number, $N$, of degrees of freedom. Hence, the changes due to noncommutativity can be accounted for, by keeping equations (\ref{Eq2.1}), (\ref{Eq2.1a}) and (\ref{Eq2.2}) unchanged, while equations (\ref{Eq2.4}) and (\ref{Eq2.5}) are replaced by:
\be\label{Eq2.8.1}
T= {2Mc^2 \over k_B N} g(N),
\ee
where $g(N)$ is some real function of $N$.

All steps considered, we obtain:
\be\label{Eq2.8.2}
F_{NC}= {GMm \over r^2} {\hbar_{eff} \over \hbar} g(N),
\ee
Let us assume that the function $g(N)$ can be written in terms of powers of $N$, i.e. $g(N) \propto N^{\beta}$, where $\beta$ is an integer constant. The value of $\beta$ can be found from the following arguments. If $\beta<0$, this will lead to a law $F_{NC} \propto r^{- 2 (1 + | \beta|)}$, which is an unacceptable modification of the ISL. However, it is easy to see that $\beta > 0$ gives origin to equally unacceptably large corrections. Indeed, the first order correction in $N$ gives origin to a constant acceleration, as in the case of the so-called Pioneer anomaly, $a_{Pio} \simeq 10^{-9}~m/s^2$. Recent calculations strongly indicate that this anomalous acceleration is due to onboard thermal effects \cite{BFGP08,BFGP11}. In any case, the first order entropic correction due to $N$ is huge, about $10^{93} a_{Pio}$, if $M$ is Earth's mass. This is of course untenable. Therefore we set $\beta=0$ and assume that the proportionality constant to be of O(1). We are thus left with Eq. (\ref{Eq2.10}).

We still have to estimate the effective Planck constant. This can be regarded (up to a multiplicative constant) as the unit phase-space cell. For a two dimensional system the minimal phase-space cell has volume $\hbar^2$. In the present situation, we want to minimize the volume functional:
\be\label{Eq2.11}
V(\Delta x_1, \Delta p_1, \Delta x_2,\Delta p_2)=\Delta x_1 \Delta p_1 \Delta x_2 \Delta p_2~,
\ee
subject to the following constraints
\be\label{Eq2.12}
\begin{array}{l l}
\Delta x_1 \Delta p_1 \ge \hbar/2, & \Delta x_1 \Delta x_2 \ge \theta_{12}/2~,\\
& \\
\Delta x_2 \Delta p_2 \ge \hbar/2, & \Delta p_1 \Delta p_2 \ge \eta_{12}/2~.
\end{array}
\ee
By the very nature of these constraints, this optimization problem should, in principle, be solved by application of the Karush-Kuhn-Tucker Theorem \cite{Jahn}. However, this is actually an ill-posed problem since there is no minimizer of the volume functional which satisfies the previous set of constraints. In fact, this is not surprising. It is well-known that the quantum uncertainty relations (\ref{Eq2.12}) cannot be all saturated simultaneously. More precisely, a quantum state can saturate at most one of these conditions \cite{Bolonek,Kosinski}.

There is thus, no minimal volume of the phase-space cell. If we choose to minimize the product $\Delta x_1 \Delta x_2$, which corresponds to $\theta_{12}/2$, then the volume functional will be minimal for $\Delta p_1 \Delta p_2=\eta_{12}/2$. Altogether,
\be\label{Eq2.13}
V(\Delta x_1, \Delta p_1, \Delta x_2,\Delta p_2) \ge {\theta_{12}\eta_{12}\over4}~.
\ee
Similarly, if we choose to minimize the product $\Delta x_1 \Delta p_1$, corresponding to $\hbar/2$, then the volume functional will be minimal for $\Delta x_2 \Delta p_2 = \hbar /2$, and thus
\be\label{Eq2.14}
V(\Delta x_1, \Delta p_1, \Delta x_2,\Delta p_2) \ge {\hbar^2\over4}
\ee
A reasonable assumption in order to comply with Eqs. (\ref{Eq2.13}) and (\ref{Eq2.14}) is
\be\label{Eq2.15}
V(\Delta x_1, \Delta p_1, \Delta x_2,\Delta p_2) \ge {\hbar^2\over4} + {\theta_{12}\eta_{12}\over4} \equiv {\hbar_{eff}^2\over4}~,
\ee
with
\be\label{Eq2.16}
\hbar_{eff}= \hbar \left(1+ {\theta_{12}\eta_{12}\over\hbar^2} \right)^{{1\over2}}~.
\ee
Assuming $\theta\eta / \hbar^2 << 1$, we finally obtain
\be\label{Eq2.17}
F_{NC}= {GMm\over r^2} \left(1+ {\theta_{12}\eta_{12}\over2 \hbar^2} \right)~.
\ee
Of course, the same reasoning could be applied for the other components of position and momentum.

This result corresponds to the most important phase-space noncommutative correction to Newton's ISL as obtained in Ref. \cite{Verlinde}. Notice that our result differs from the one of Ref. \cite{Nicolini} in that our correction depends explicitly on the noncommutativity of both configuration and momentum variables and it could be matched with the one of Ref. \cite{Nicolini} only if noncommutativity in the momentum variables were completely disregarded from the very beginning. Furthermore, the two results differ yet in another fundamental point, namely that the modification proposed in Ref. \cite{Nicolini} assumes a minimal length and it yields an isotropic correction to Newton's ISL, which has a dependence on the radial coordinate. Our result on the other hand, arises from considerations about a minimal volume of the unitary phase-space cell and yields an anisotropic correction to Newton's law, which implies a violation of the Equivalence Principle. Thus, we conclude that it 
 is not possible to recover the results of Ref. \cite{Nicolini} from our formulation. 

Of course, we could wonder whether the obtained correction could instead be regarded as putative corrections to the fundamental constants at play, namely Newton's constant, speed of light and Boltzmann's constant. The possibility of an anisotropic correction to Newton's constant would give rise to a violation of the Equivalence Principle as we discuss in the next section. In what concerns the other two fundamental constants, it should be realized that they completely cancel out throughout the procedure of obtaining Newton's ISL and hence, at least in principle, no conclusions can be drawn about noncommutative corrections to them. In any case, it should be realized that bounds on corrections on the speed of light are quite stringent as they imply the breaking of Lorentz invariance (see e.g. Refs. \cite{Bertolami00,BPT06}), and no credible evidence about corrections to Boltzmann's constant have ever been reported given that they would imply some failure of the extremely well es
 tablished, and presumably universal, principles of thermodynamics.

\section{Equivalence Principle and Bounds on the noncommutative parameters}

We have shown in the last section that the entropic force $F$ can acquire a correction due to noncommutativity that is not isotropic. This putative anisotropy implies that nocommutativity will affect the gravitational fall of masses on different directions. The expected  relative differential acceleration is given, if one considers, say fall along the plane $(1,2)$ with acceleration, $a_1$, and say along the plane $(2,3)$ with acceleration $a_2$:
\be\label{Eq3.1}
{\Delta a\over a} \equiv 2\left({a_1-a_2 \over a_1+a_2}\right) \simeq {1\over 4\hbar^2}(\theta_{12}\eta_{12} - \theta_{23}\eta_{23})~.
\ee
Therefore, at it stands, this result sets a bound on the anisotropy of noncommutativity. If however, we further assume that $\theta_{12}\eta_{12} \simeq O(1) \theta_{23}\eta_{23} \equiv \theta\eta$, then the Equivalence Principle, whose most stringent experimental limit is given by $\Delta a/a \lsim10^{-13}$ \cite{Adelberger}, can be used to set a bound on the dimensionless quantity $\theta\eta / \hbar^2$. Thus, we obtain\footnote{Notice that, as previously referred, the dependence of the ISL of $g(N)$ is specified up to a constant of $O(1)$. This constant can be absorbed by a redefinition of Newton's constant, which does not affect at all our result as this depends fundamentally on the anisotropic nature of the  encountered corrections to Newton's ISL.}:
\be
{\theta\eta\over \hbar^2} \lsim O(1) \times 10^{-13}~.
\label{Eq3.2}
\ee
This bound states that noncommutative effects can be 10 orders of magnitude greater than discussed in the context of the noncommutative gravitational quantum well \cite{Bertolami1}. Moreover, it does not presuppose any value for the $\theta$ parameter. This bound implies that if, for instance, $\sqrt{\theta} \lsim (10~TeV)^{-1}$ as inferred from the induced Lorentz invariance in the electromagnetic sector of the Standard Model extension due to configuration space noncommutativity \cite{Carroll}, then $\sqrt{\eta} \gsim 10^{-4}~GeV$. The latter bound is unsatisfactory as it implies that noncommutative effects should already have been observed. If however, one assumes that $\sqrt{\theta} \gsim M_P^{-1}$, i.e. that the characteristic scale of $\theta$ is essentially the scale of quantum gravity effects (the Planck mass, $M_P=L_P ^{-1}$), then $\sqrt{\eta} \gsim 10^{-6}~M_P$. This is an interesting intermediary scale. At this point, further advancement is only possible with new o
 bservational results, either via the direct identification of noncommutative effects or through more stringent bounds on the validity of the Equivalence Principle (see discussion in the next section).

\section{Discussion and Conclusions}

In this work we have generalized the entropic derivation of Newton's ISL to accommodate phase-space noncommutative effects. The modifications were argued to be tantamount to an effective Planck constant. This allowed us to get Newton's ISL law corrected by a noncommutative contribution. Our result arises from the uncertainty relations ensued by the phase space noncommutative algebra and considerations about a minimal size cell in the phase space, and depend fundamentally on the anisotropy of the noncommutative parameters. This contrasts, for instance with the result of Ref. \cite{Nicolini}, where only configuration space noncommutatitity is considered and a minimal length is assumed, and that leads to an isotropic correction to Newton's law. The anisotropic nature of the noncommutative correction that we have obtained implies a violation of the Equivalence Principle, whose experimental bound yields a quite interesting bound on the noncommutativity anisotropy, and under assump
 tion that the product of the configuration space and the momentum space noncommutative parameters at different directions differ by an O(1) constant, that $\theta\eta / \hbar^2 \lsim O(1) \times 10^{-13}$.

Our reasoning and results rely on (i) the arguments of Ref. \cite{Verlinde} in what concerns the emergence of macroscopic space and gravity and their essential connection with the Holographic Principle; and (ii) the assumption that phase-space noncommutative features should be incorporated. However, one could wonder whether the standing of Eq. (\ref{Eq2.1}) is inconsistent with the experimental realization of the gravitational quantum well (GQW) using ultra-cold neutrons in the GRANIT experiment \cite{granitN}. With respect to this criticism, we argue that there is no reason to assume that the typical length scale of this experiment, $l_{GQW} \simeq O(1) \mu m$, is the one associated with entropy change, $\Delta x$, in Eq. (\ref{Eq2.1}). Indeed, assuming that space is an emerging feature from a coarse graining procedure, the question is: what is the graining scale? It seems natural to assume that the former is such that it is much greater than the fundamental area from which 
 one counts the number of bits, namely $L_P^2$, the squared Planck's length. Thus, it is natural to expect the following hierarchy of scales: $L_p<< \lambda_c << l_{GQW}$. And similarly, we expect that $\Delta x \lsim \lambda_c$ for the displacement associated with the Bekenstein bound. Actually, in our understanding, what the experimental realization of the GQW really establishes is a ``gravitational" drawing line between quantum and classical behaviour given the dependence of its energy
spectrum on the mass of the testing particle. This means that a particle can be regarded as quantum from the gravitational point of view only if its size is smaller than the spacing between the energy levels of the corresponding GQW \cite{Bertolami06}.

Another surprising implication of our result concerns the connection with the observed value of the cosmological constant. Indeed, it has been argued that a putative breaking of the Equivalence Principle at about $\Delta_{EP} \simeq10^{-14}$ level implies that the vaccum energy is related with the observed discrepancy between the observed value of the cosmological constant and the expected value from the electroweak symmetry breaking in the Standard Model \cite{Bertolami10}:
\be
{\Lambda_{Obs.} \over \Lambda_{SM}} = {\Delta^4_{EP}}~,
\label{Eq4.1} 
\ee
and thus
\be
{\Lambda_{Obs.}\over \Lambda_{SM}} \simeq \left({\theta\eta\over \hbar^2}\right)^4 ~,
\label{Eq4.2}
\ee
a quite interesting relationship. Notice that the bound Eq. (\ref{Eq3.2}) is fairly close the most stringent one that can be achieved from the above considerations.

Thus, the emergence of gravity through thermodynamical arguments, besides its own intrinsic pertinence, provides a suggestive connection between the parameters of phase-space noncommutativity with the observed value of the cosmological constant.
Of course, this relationship holds as far as the Equivalence Principle is found to be violated at $\Delta_{EP} \simeq10^{-14}$ level, just an order of magnitude beyond the current bound.

\subsection*{Acknowledgments}

\vspace{0.3cm}

\noindent The work of CB is supported by Funda\c{c}\~{a}o para a Ci\^{e}ncia e a Tecnologia (FCT) under the grant SFRH/BPD/62861/2009. The work of OB is partially supported be FCT project PTDC/FIS/111362/2009. The work of NCD and JNP is partly supported by the grants PTDC/MAT/69635/2006 and PTDC/MAT/099880/2008 of FCT.


\end{document}